\documentclass[10pt,twocolumn,twosidet]{IEEEtran}

\usepackage{amsmath}
\usepackage{subfigure}
\usepackage{caption}
\usepackage{url}
\usepackage{amssymb}
\usepackage{theorem}

\usepackage{tikz}
\usepackage{tkz-berge}
\usepackage{pgfplots}
\usetikzlibrary{plotmarks}

\usepackage{psfrag}

\newtheorem{Prop}{Proposition}[section]
\newtheorem{Def}{Definition}[section]

\ifCLASSINFOpdf
\else
\fi

\begin{document}
%
\title{Adaptive non-Zero Mean Gaussian Detection and Application to Hyperspectral Imaging}
\author{Joana Frontera-Pons,~\IEEEmembership{Student Member,~IEEE,}
            Fr\'ed\'eric Pascal,~\IEEEmembership{Member,~IEEE,}
            and~Jean-Philippe Ovarlez,~\IEEEmembership{Member,~IEEE}%
        \thanks{J. Frontera-Pons is with SONDRA, Supelec, Plateau du Moulon, 3 rue Joliot-Curie, F-91190 Gif-sur-Yvette, France (e-mail: joana.fronterapons@supelec.fr)}
\thanks{F. Pascal is with SONDRA, Supelec, Plateau du Moulon, 3 rue Joliot-Curie, F-91190 Gif-sur-Yvette, France (e-mail: frederic.pascal@supelec.fr)}
\thanks{J.-P. Ovarlez is with ONERA, DEMR/TSI, Chemin de la Huni\`ere, F-91120 Palaiseau, France (e-mail: jean-philippe.ovarlez@onera.fr)}}


\maketitle


\begin{abstract}
Classical target detection schemes are usually obtained deriving the likelihood ratio under Gaussian hypothesis and replacing the unknown background parameters by their estimates. In most applications, interference signals are assumed to be Gaussian with zero mean or with a known mean vector that can be removed and with unknown covariance matrix. When mean vector is unknown, it has to be jointly estimated with the covariance matrix, as it is the case for instance in hyperspectral imaging. In this paper, the adaptive versions of the classical Matched Filter and the Normalized Matched Filter, as well as two versions of the Kelly detector are first derived and then are analyzed for the case when the mean vector of the background is unknown. More precisely, theoretical closed-form expressions for false-alarm regulation are derived and the Constant False Alarm Rate property is pursued to allow the detector to be independent of nuisance parameters. Finally, the theoretical contribution is validated through simulations and on real hyperspectral scenes. 
\end{abstract}


%
\IEEEpeerreviewmaketitle

\section{Introduction}

\IEEEPARstart{H}{yperspectral} imaging (HSI) extends from the fact that for any given material, the amount of radiation emitted varies with wavelength. HSI sensors measure the radiance of the materials within each pixel area at a very large number of contiguous spectral bands and provide image data containing both spatial and spectral information (see for more details \cite{chang2003hyperspectral} and reference therein). Hyperspectral processing involves various applications such as unmixing, classification, detection, dimensionality reduction, ... Among them, hyperspectral detection is an active research topic that has led to many publications e.g. \cite{manolakis2002detection, stein2002anomaly, chang2002anomaly, kwon2006kernel}. More precisely, hyperspectral target detection methods are commonly used to detect targets embedded in background and that generally cannot be solved by spatial resolution \cite{matteoli2010tutorial}. Furthermore, Detection Theory \cite{kay1998fundamentals} arises in many different military and civilian applications and has been widely investigated in several signal processing domains such as radar, sonar, communications, see \cite{gini2001selected} for the different references. There are two different methodologies for target detection purposes in the HSI literature \cite{manolakis2003hyperspectral}: Anomaly Detection \cite{stein2002anomaly, chang2002anomaly} and Target Detection \cite{manolakis2002detection}.\\
In many practical situations, there is not enough information about the target to detect, thus  \emph{Anomaly Detection} methods are widely used. The most widespread detector, the RX detector \cite{reed1990adaptive} is based on the Mahalanobis distance \cite{mahalanobis1936generalized}. This detector and most of its variants search for pixels in the image with spectral characteristics that differ from the background. On the other hand, when the spectral signature of the desired target is known, it can be used as steering vector in \emph{Target Detection} techniques \cite{manolakis2003hyperspectral}.\\

Interestingly, target detection methods have been extensively developed and analyzed in the signal processing and radar processing \cite{gini2001selected, kelly1986adaptive, Kraut01, robey1992cfar}. In all these works as well as in several signal processing applications, signals are assumed to be Gaussian with zero mean or with a known mean vector (MV) that can be removed. In such context, Statistical Detection Theory \cite{kay1998fundamentals} has led to several well-known algorithms, for instance the Matched Filter (MF) and its adaptive versions, the Kelly detector \cite{kelly1986adaptive} or the Adaptive Normalized Matched Filter \cite{kraut1999cfar}. Other interesting approaches based on subspace projection methods have been derived and analyzed in \cite{Kraut01}. However, when the mean vector of the noise background is unknown, these techniques are no longer adapted and improved methods have to be derived by taking into account the mean vector estimation. For this purpose, some preliminary results have been given in \cite{frontera2013false}. One of the contributions of this work is to extend and generalize these original results.\\

More precisely, this work deals with the classical Adaptive Matched Filter (AMF), the Kelly detection test and the Adaptive Normalized Matched Filter (ANMF). These detectors have been derived under Gaussian assumptions and benefit from great popularity in HSI target detection literature, see e.g. \cite{manolakis2009there,manolakis2013remarkable}. To evaluate the detector performance, the classical process, according to the Neyman-Pearson criterion is first to regulate the false-alarm, by setting a detection threshold for a given probability of false-alarm (PFA). Since the PFA is the cumulative distribution function (CDF) of the detection test, this process is equivalent to the derivation of the detection test distribution. Then, the probability of detection is evaluated for different Signal-to-Noise Ratios (SNR). Therefore, keeping the false-alarm rate constant (CFAR) is essential to set a proper detection threshold \cite{Gini02,Conte02a}. The aim is to build a CFAR detector which provides detection thresholds that are relatively immune to noise and background variation, and allow target detection with a constant false-alarm rate. The theoretical analysis of CFAR methods for adaptive detectors is a challenging problem since in adaptive schemes, the statistical distribution of the detectors is not always available in a closed-form expression.\\

The theoretical contributions of this paper are twofold. First, we derive the expression of each adaptive detector under the Gaussian assumption where both then mean vector and the covariance matrix (CM) are assumed to be unknown. Then, the exact derivation of the distribution of each proposed detection scheme under null hypothesis, i.e. when no target is supposed to be present, is provided. Thus, through Gaussian assumption, closed-form expressions for the false-alarm regulation are obtained, which allow to theoretically set the detection threshold for a given PFA.\\

One the other hand, one difficulty for the background detection statistic is to assume a tractable model or at least to account for robustness to deviation from the assumed theoretical model in the detection scheme.  Since Gaussian assumption is not always fulfilled for real hyperspectral data, alternative robust estimation techniques are proposed in \cite{frontera2012a}. However, it is essential to notice that the derivations for many results in robust detection contexts strongly rely on the results obtained in the Gaussian context. For instance, this is the case of \cite{Pascal06} in which the derivation of a robust detector distribution is based on its Gaussian counterpart.\\

This paper is organized as follows. Section \ref{sec2} introduces the required background on classical detection techniques as well as the obtention of the adaptive detectors for both unknown MV and CM. Then, Section \ref{sec3} provides the main theoretical contributions of the paper by deriving the exact "PFA-threshold" relationship for the AMF, the "plug-in" Kelly detector and the ANMF under Gaussian assumption while a generalized version of the Kelly detector is derived. Finally, in Section \ref{sec4}, the theoretical analyses are validated through Monte-Carlo simulations and real HS data are processed to, first, extract homogeneous, let's say Gaussian, data and then, highlight  the agreement with the proposed theoretical results.  Conclusions and perspectives are drawn in Section \ref{sec5}.\\

In the following, vectors (resp. matrices) are denoted  by bold-faced lowercase letters (resp. uppercase letters). ${}^T$ and ${}^H$ respectively represent the transpose and the Hermitian operators. $|\mathbf{A}|$ represents the determinant of the matrix $\mathbf{A}$ and $\text{Tr}(\mathbf{A})$ its trace. $j$ is used to denote the unit imaginary number. $\sim$ means "distributed as". $\Gamma(\cdot)$ denotes the gamma function. Eventually, $\Re\{\mathbf{x}\}$ represents the real part of the complex vector $\mathbf{x}$. 

\section{Background and adaptive detectors derivation}
\label{sec2}
After providing the general background in non-zero mean Gaussian detection, this section is devoted to the derivation of the expression of the adaptive detectors.

%
The problem of the detecting a known signal $\mathbf s$ corrupted by an additive noise $\mathbf b$ in a $m$-dimensional complex vector \textbf{x} can be stated as a the following binary hypothesis test:
\[
\begin{cases}
\mathcal{H}_0 : \mathbf{x} = \mathbf{b} &  \\
\mathcal{H}_1: \mathbf{x} = \mathbf{s} + \mathbf{b} &,  
\end{cases}
\]
and the signal $\mathbf{s}$ can be written in the form $\alpha \mathbf{p}$, where $\alpha$ is an unknown complex scalar amplitude, and $\mathbf{p}$ is the steering vector describing the signal which is sought. Since the background statistics, i.e. the MV and the CM, are assumed to be unknown, they have to be estimated from $\mathbf{x}_1,...\mathbf{x}_N \sim \mathcal{CN} (\boldsymbol{\mu, \boldsymbol{\Sigma}})$ a sequence of $N$ IID signal-free secondary data. Then, the adaptive detector is obtained by replacing the unknown parameters by their estimates. In practice, an estimate may be obtained from the range cells surrounding the cell under test, which play the role of the $N$ IID signal-free secondary data. The sample size $N$ has to be chosen large enough to ensure the invertibility of the covariance matrix and small enough to justify both spectral homogeneity (stationarity) and spatial homogeneity. The use of a sliding mask provides a more realistic scenario than when estimating the parameters using all the pixels in the image. Let us know recall the detectors under interest in this work

\subsection{Adaptive Matched Filter}
The MF detector is the optimal linear filter for maximizing the SNR in the presence of additive Gaussian noise with known parameters \cite{kay1998fundamentals}.
Hence, the signal model can be written as:
\begin{equation}
\begin{cases}
\mathcal{H}_0 :  &\,  \mathbf{x = b} \,\sim \mathcal{CN} (\boldsymbol{\mu},\boldsymbol{\Sigma}) \\
\mathcal{H}_1 :  &\,   \mathbf{x} = \alpha \mathbf{p + b} \sim \mathcal{CN} (\alpha \mathbf{p} + \boldsymbol{\mu},\boldsymbol{\Sigma}).
\end{cases}
\label{MF_signal}
\end{equation}
The Likelihood Ratio (LR) is given by:
\begin{equation*}
L(\alpha) = \cfrac{f(\mathbf{x}|\mathcal{H}_1)}{f(\mathbf{x}|\mathcal{H}_0)} \underset {H_0} {\overset {H_1} {\gtrless}} \lambda \, 
\end{equation*}
or according to the signal model:
\begin{equation}
L(\alpha) = \cfrac{\exp [-\left(\mathbf{x}-(\alpha \mathbf{p} + \boldsymbol{\mu})\right)^H\boldsymbol{\Sigma}^{-1} \left(\mathbf{x}-(\alpha \mathbf{p} + \boldsymbol{\mu})\right)]}{\exp [ -(\mathbf{x}-\boldsymbol{\mu})^H\boldsymbol{\Sigma}^{-1} (\mathbf{x}- \boldsymbol{\mu})]} \underset {H_0} {\overset {H_1} {\gtrless}} \lambda . 
\label{LR_MF}
\end{equation}
Since the complex amplitude is unknown, it has to be estimated from the observation vector $\mathbf{x}$ and the background parameters according to:
\begin{equation}
\alpha = \cfrac{\Re\{\mathbf{p}^H\boldsymbol{\Sigma}^{-1}(\mathbf{x}-\boldsymbol{\mu})\}}{\mathbf{p}^H\boldsymbol{\Sigma}^{-1}\mathbf{p}}.
\label{amplitude}
\end{equation}
Replacing this value in \eqref{LR_MF} and after some manipulations, the resulting MF detection scheme is:
\begin{equation}
\Lambda_{MF} = \cfrac {|\mathbf{p}^H \, {\boldsymbol{\Sigma}}^{-1} \, (\mathbf{x} - {\boldsymbol{\mu}})|^2}{(\mathbf{p}^H \, {\boldsymbol{\Sigma}}^{-1} \, \mathbf{p})} \underset {H_0} {\overset {H_1} {\gtrless}} \lambda\, .
\label{MF}
\end{equation}
Note that it differs from the classical MF by the term $\boldsymbol{\mu}$, the background mean, but without any consequence since $\mathbf x -\boldsymbol{\mu} \sim \mathcal{CN}(\mathbf{0},\boldsymbol{\Sigma})$. Moreover, the "PFA-threshold" relationship is given by \cite{kay1998fundamentals}:
\[
 PFA_{MF} = \exp{(-\lambda)} .
\]
The AMF, denoted $\Lambda_{AMF}^{(N)} {}_{\hat{\boldsymbol{\Sigma}}}$ to underline the dependency with $N$, is usually built replacing the covariance matrix $\boldsymbol{\Sigma}$ by its estimate $\hat{\boldsymbol{\Sigma}}$ obtained from the $N$ secondary data. The mean vector is generally supposed to be known. Thus, the adaptive version becomes:
\begin{equation*}
\Lambda_{AMF}^{(N)} {}_{\hat{\boldsymbol{\Sigma}}} = \cfrac {|\mathbf{p}^H \,{\hat{\boldsymbol{\Sigma}}}^{-1} \, (\mathbf{x} - {{\boldsymbol{\mu}}})|^2}{(\mathbf{p}^H \,{\hat{\boldsymbol{\Sigma}}}^{-1} \, \mathbf{p})} \underset {H_0} {\overset {H_1} {\gtrless}} \lambda \, .
\end{equation*}
Then, the theoretical "PFA-threshold" relationship is given by \cite{robey1992cfar}  for $\hat{\boldsymbol{\Sigma}}=\hat{\boldsymbol{\Sigma}}_{SCM}$:
\begin{equation}
PFA_{AMF}{}_{\hat{\boldsymbol{\Sigma}}} = {}_2F_1\left( N-m+1, \,N-m+2;\, N+1;  \, - \frac{\lambda} {N} \right) ,
\label{pfa_AMF}
\end{equation}
where ${}_2F_1(\cdot)$ is the hypergeometric function \cite{abramowitz1964handbook}
defined as,
\begin{equation*}
{}_2 F_1(a,b;c;z) = \cfrac{\Gamma(c)}{\Gamma(b)\Gamma(c-b)} \int_0^1 \cfrac{t^{b-1}(1-t)^{c-b-1}}{(1-tz)^a} dt\, .
\end{equation*}
This detector holds the CFAR properties in the sense that its false alarm expression only depends on the dimension of the vector $m$ and the number of secondary data used for the estimation $N$. Note that it is also independent of the noise covariance matrix $\boldsymbol{\Sigma}$, therefore the detector is said to be CFAR-matrix. However, its performance strongly relies on the good fit of the Gaussian model  and the false alarm rate is highly increased when normal assumption is not verified.\\

\subsection{Adaptive Kelly detector}
The Kelly detector was derived in \cite{kelly1986adaptive}. It is based on the Generalized Likelihood Ratio Test (GLRT) assuming Gaussian distribution and the same signal model than the AMF in \eqref{MF_signal}. In this case, only the covariance matrix $\boldsymbol{\Sigma}$ is unknown, the mean vector is assumed to be known. Thus, the joint probability density function (p.d.f.) of the the $N$ secondary data and the observation vector $\mathbf{x}$ under the two hypotheses $\mathcal{H}_i$ can be written as:
\begin{equation}\label{LF-Kelly}
f_i(\mathbf{x}) = \left( \cfrac{1}{\pi^m |\boldsymbol{\Sigma}|} \exp [-\text{Tr} (\boldsymbol{\Sigma}^{-1} \mathbf{T}_i)]\right)^{N+1} \, ,
\end{equation}
where $\mathbf{T}_i$ is the composite sample covariance matrix constructed from both the secondary data and observation vector:
\begin{align*}
\mathbf{T}_0 &= \cfrac{1}{N+1}\left(  (\mathbf{x} - {\boldsymbol{\mu}})(\mathbf{x} - {\boldsymbol{\mu}})^H + \hat{\mathbf{W}}  \right)\\
\mathbf{T}_1 &= \cfrac{1}{N+1}\left(  (\mathbf{x} - (\alpha \mathbf{p}+{\boldsymbol{\mu}})) (\mathbf{x} - (\alpha \mathbf{p}+{\boldsymbol{\mu}}))^H
+\hat{\mathbf{W}} \right)
\end{align*}
and $\hat{\mathbf{W}} = N \, \hat{\boldsymbol{\Sigma}}_{SCM}$, where $\hat{\boldsymbol{\Sigma}}_{SCM}$ represents the well-known Sample Covariance Matrix (SCM) recalled in Appendix \ref{app1}.  
Then, by maximizing the p.d.f under both hypotheses and by maximizing the LR with respect to (w.r.t) the complex, and after some manipulations, the resulting adaptive Kelly detector scheme takes the following form:
\begin{align}\label{LR-Kelly}
&\Lambda_{Kelly}^{(N)} {}_{\hat{\boldsymbol{\Sigma}}} \\ &= \cfrac {|\mathbf{p}^H \, \hat{\boldsymbol{\Sigma}}_{SCM}^{-1} \, (\mathbf{x} - {{\boldsymbol{\mu}}})|^2}{\left(\mathbf{p}^H \, \hat{\boldsymbol{\Sigma}}_{SCM}^{-1}\mathbf{p}\right) \, \left(  N + (\mathbf{x} - {{\boldsymbol{\mu}}})^H \, \hat{\boldsymbol{\Sigma}}_{SCM}^{-1} \, (\mathbf{x} - {{\boldsymbol{\mu}}})\right)} \underset {H_0} {\overset {H_1} {\gtrless}} \lambda\,,
\end{align}
where
$
\lambda = 1-\eta^{-\frac{1}{N+1}}.
$ 
As shown in \cite{kelly1986adaptive}, the PFA for the Kelly test is given by:
\begin{equation}
PFA_{Kelly} = (1 - \lambda)^{N - m + 1}\,.
\end{equation}
The Kelly detector is a CFAR test, in which the PFA is independent of the true covariance matrix. However, it has no known optimality property in the sense of maximizing the probability of detection for a given probability of false alarm. 
The AMF and the Kelly detector are based on the same assumptions about the nature of the observations. It is therefore interesting to compare their detection performance for a  given PFA. Note that for large values of $N$ the performances are substantially the same.\\

\subsection{Adaptive Normalized Matched Filter}
The Normalized Matched Filter (NMF) is obtained when considering that the covariance matrix is different under the two hypotheses. That is to say that the clutter has the same covariance structure but different variance.
\begin{equation*}
\begin{cases}
\mathcal{H}_0 :  &\,  \mathbf{x = b} \,\sim \mathcal{CN} (\boldsymbol{\mu},\sigma_0^2\boldsymbol{\Sigma}) \\
\mathcal{H}_1 :  &\,   \mathbf{x} = \alpha \mathbf{p + b} \sim \mathcal{CN} (\alpha \mathbf{p} + \boldsymbol{\mu},\sigma_1^2\boldsymbol{\Sigma}).
\end{cases}
\end{equation*}
Thus, the ML estimates of $\sigma_j^2$ are easily derived from $\hat{\sigma}^2_j = \operatorname{arg\,max}_\sigma^2 \{ f(\mathbf{x}|\sigma,\mathcal{H}_j)\} , (j = 0,1)$ and assuming normal distribution, one has:
\begin{align*}
\hat{\sigma}_0^2 =& \cfrac{1}{2m} (\mathbf{x}-\boldsymbol{\mu})^H \boldsymbol{\Sigma}^{-1}(\mathbf{x}-\boldsymbol{\mu})\\
\hat{\sigma}_1^2 =& \cfrac{1}{2m} \left(\mathbf{x}- ( \alpha \mathbf{p} + \boldsymbol{\mu})\right)^H \boldsymbol{\Sigma}^{-1} \left(\mathbf{x}- ( \alpha \mathbf{p} + \boldsymbol{\mu})\right)
\end{align*}
After replacing complex amplitude $\alpha$ by its estimate \eqref{amplitude} when building the LR and after some manipulations, one obtains \cite{scharf1994matched}: 
\begin{equation}\label{LR-ANMF}
\Lambda_{NMF} = \cfrac {|\mathbf{p}^H \, {\boldsymbol{\Sigma}}^{-1} \, (\mathbf{x} - {{\boldsymbol{\mu}}})|^2}{(\mathbf{p}^H \, {\boldsymbol{\Sigma}}^{-1}\mathbf{p}) \, \left( (\mathbf{x} - {{\boldsymbol{\mu}}})^H \, {\boldsymbol{\Sigma}}^{-1} \, (\mathbf{x} - {{\boldsymbol{\mu}}})\right)} \underset {H_0} {\overset {H_1} {\gtrless}} \lambda,
\end{equation}
where
$\lambda = 1-\eta^{-\frac{1}{m}}$ and for which one has \cite{scharf1994matched}:
\[
PFA_{NMF} = (1-\lambda)^{(m-1)} \, .
\]
The ANMF is generally obtained when the unknown noise covariance matrix is replaced by an estimate \cite{Kraut01}: 
\begin{equation*}
\Lambda_{ANMF}^{(N)} {}_{\hat{\boldsymbol{\Sigma}}} = \cfrac {|\mathbf{p}^H \, \hat{\boldsymbol{\Sigma}}^{-1} \,  (\mathbf{x} - {\boldsymbol{\mu}})|^2}{\left(\mathbf{p}^H \, {\hat{\boldsymbol{\Sigma}}}^{-1}\mathbf{p}\right) \, \left( (\mathbf{x} - {{\boldsymbol{\mu}}})^H \,{\hat{\boldsymbol{\Sigma}}}^{-1} \, (\mathbf{x} - {{\boldsymbol{\mu}}})\right)} \underset {H_0} {\overset {H_1} {\gtrless}} \lambda.
\end{equation*}
And the PFA follows \cite{Kraut01} for $\hat{\boldsymbol{\Sigma}}=\hat{\boldsymbol{\Sigma}}_{SCM}$:
\begin{equation}
PFA_{ANMF}{}_{\hat{\boldsymbol{\Sigma}}} = (1-\lambda)^{a-1} \, _2 F_1(a,a-1;b-1;\lambda)\,,
\label{pfa_ANMF}
\end{equation}
where $a = N-m+2$ and $b = N+2$.

\section{Main Results}
\label{sec3}

In this section, let us now assume that the mean parameter is unknown as it is the case for instance in HSI and let us derive the new corresponding detection schemes. Then, using standard calculus on Wishart distributions, recapped in Appendix \ref{app2}, the distributions of each detection test is provided.

\subsection{Adaptive Matched Filter Detector}
When both covariance matrix and mean vector are unknown, they are replaced by their estimates from the secondary data in \eqref{MF} leading to the AMF detector of the following form:
\begin{equation*}
\Lambda_{AMF}^{(N)} {}_{\hat{\boldsymbol{\Sigma}},\hat{\boldsymbol{\mu}}}  = \cfrac {|\mathbf{p}^H \, {\hat{\boldsymbol{\Sigma}}}^{-1} \, (\mathbf{x} - {\hat{\boldsymbol{\mu}}})|^2}{(\mathbf{p}^H \, {\hat{\boldsymbol{\Sigma}}}^{-1} \, \mathbf{p})} \underset {H_0} {\overset {H_1} {\gtrless}} \lambda,
\end{equation*}
where the notation $\Lambda_{AMF}^{(N)} {}_{\hat{\boldsymbol{\Sigma}},\hat{\boldsymbol{\mu}}}$ is used to stretch now the dependency on the estimated mean vector $\hat{\boldsymbol{\mu}}$. The distribution of this detection test is given in the next Proposition, through its PFA.
\begin{Prop}\label{Prop-Pfa_AMF_mean}
Under Gaussian assumptions, the theoretical relationship between the $PFA$ and the threshold is given by
\begin{equation} \label{Pfa_AMF_mean}
PFA_{AMF}{}_{\hat{\boldsymbol{\Sigma}},\hat{\boldsymbol{\mu}}} = {}_2F_1\left( N-m, \,N-m+1;\, N;  \, - \frac{\lambda'} {N-1} \right)\, ,
\end{equation}
where $\lambda' = {\frac{(N-1)}{(N+1)}} \lambda$, $\hat{\boldsymbol{\Sigma}}=\hat{\boldsymbol{\Sigma}}_{SCM}$ and $\hat{\boldsymbol{\mu}}=\hat{\boldsymbol{\mu}}_{SMV}$, recapped in Appendix \ref{app1}.
\end{Prop}

Before turning into the proof, let us comment on this result. 
\begin{itemize}
\item Interestingly, this detector also holds the CFAR property in the sense that its false-alarm expression depends only on the dimension $m$ and on the number of secondary data $N$, but not on the noise parameters $\boldsymbol{\mu}$ and $\boldsymbol{\Sigma}$. Note that the only effect of estimating the mean is the loss of one degree of freedom and the modification of the threshold compared to eq. \eqref{pfa_AMF}. Obviously, the impact of these modification decreases as the number of secondary data $N$ used to estimate the unknown parameters increases.\\
\item Moreover, the result has been obtained when using the MLEs of the unknown parameters but the proof can be easily extended to other covariance matrix estimators such as $\displaystyle\hat{\boldsymbol{\Sigma}}=\cfrac{1}{N-1}  \sum_{i=1}^N (\mathbf{x}_i-\hat{\boldsymbol{\mu}}) (\mathbf{x}_i-\hat{\boldsymbol{\mu}})^H$ which the unbiased covariance matrix estimate or $\displaystyle\hat{\boldsymbol{\Sigma}}=\cfrac{1}{N+1}  \sum_{i=1}^N (\mathbf{x}_i-\hat{\boldsymbol{\mu}}) (\mathbf{x}_i-\hat{\boldsymbol{\mu}})^H$.\\
\end{itemize}

\begin{IEEEproof}
For simplicity matters, the following notations are used: $\hat{\boldsymbol{\Sigma}}=\hat{\boldsymbol{\Sigma}}_{SCM}$ and $\hat{\boldsymbol{\mu}}=\hat{\boldsymbol{\mu}}_{SMV}$.\\
Since the derivation of the PFA is done under hypothesis $H_0$, let us set $\forall i=1,...,N, \mathbf{x}_i \sim \mathcal{CN} (\boldsymbol{\mu},\boldsymbol{\Sigma})$ and $\mathbf{x}\sim \mathcal{CN} (\boldsymbol{\mu},\boldsymbol{\Sigma})$, where all these vectors are independent. Now, let us denote $$\hat{\mathbf{W}}_{N-1} = \sum_{i=1}^N (\mathbf{x}_i-\hat{\boldsymbol{\mu}}) (\mathbf{x}_i-\hat{\boldsymbol{\mu}}) ^H \sim \mathcal{CW} (N-1,\boldsymbol{\Sigma}),
$$
Since $\hat{\boldsymbol{\mu}} \sim \mathcal{CN} (\boldsymbol{\mu}, \frac{1}{N}\boldsymbol{\Sigma})$, one has $\mathbf{x} - {\hat{\boldsymbol{\mu}}} \sim\mathcal{CN} (\mathbf{0}, \frac{N+1}{N}\boldsymbol{\Sigma}).$ This can be equivalently rewritten as
\[
 \sqrt{{N}/{(N+1)}} (\mathbf{x} - {\hat{\boldsymbol{\mu}}}) \sim \mathcal{CN} (\mathbf{0}, \boldsymbol{\Sigma}).
\]
Now, let us set  $\mathbf{y} = \sqrt{\frac{N}{N+1}}\left(\mathbf{x} - {\hat{\boldsymbol{\mu}}}\right)$ with $\mathbf y \sim \mathcal{CN} (\mathbf{0}, \boldsymbol{\Sigma})$. 

When computing the SCM, one has
\begin{eqnarray*}
\hat{\boldsymbol{\Sigma}}_{SCM} = &  \cfrac{1}{N}  \displaystyle\sum_{i=1}^N (\mathbf{x}_i-\hat{\boldsymbol{\mu}}) (\mathbf{x}_i-\hat{\boldsymbol{\mu}}) ^H = \cfrac{1}{N}\hat{\mathbf{W}}_{N-1}.
\end{eqnarray*}

As we jointly estimate the mean and the covariance matrix, a degree of freedom is lost, compared with the only covariance matrix estimation problem.  

Let us now consider the classical AMF test (i.e. $\boldsymbol{\mu}$ known) built from $N-1$ secondary data, rewritten in terms of $\hat{\mathbf{W}}_{N-1}$:
\[
\Lambda_{AMF}^{(N-1)} {} _{\hat{\boldsymbol{\Sigma}}} = (N-1)\, \cfrac{|\mathbf{p}^H \, {\hat{\mathbf{W}}}_{N-1}^{-1} \, \mathbf{y}|^2}{(\mathbf{p}^H \, {\hat{\mathbf{W}}}_{N-1}^{-1} \, \mathbf{p})},
\]
where $\mathbf{y}\sim \mathcal{CN} (\mathbf{0}, \boldsymbol{\Sigma})$ and whose "PFA-threshold" relationship is given by eq. \eqref{pfa_AMF} where $N$ is replaced by $N-1$.

Now, for the joint estimation problem, the AMF can be rewritten as:
\begin{align*}
\Lambda_{AMF}^{(N)}{} _{\hat{\boldsymbol{\Sigma}}, \hat{\boldsymbol{\mu}}} & = N\, \cfrac{|\mathbf{p}^H \, {\hat{\mathbf{W}}}_{N-1}^{-1} \, (\mathbf{x}-\hat{\boldsymbol{\mu}})|^2}{(\mathbf{p}^H \, {\hat{\mathbf{W}}}_{N-1}^{-1} \, \mathbf{p})}\\ & = N\,{ \cfrac{N+1}{N}}\, \cfrac{|\mathbf{p}^H \, {\hat{\mathbf{W}}}_{N-1}^{-1} \, \mathbf{y}|^2}{(\mathbf{p}^H \, {\hat{\mathbf{W}}}_{N-1}^{-1} \, \mathbf{p})} \\    &  = {\cfrac{(N+1)}{(N-1)}} \, \Lambda_{AMF}^{(N-1)}{} _{\hat{\boldsymbol{\Sigma}}}
\end{align*}
where $(\mathbf{x}-\hat{\boldsymbol{\mu}})$ has been replaced by ${\sqrt{N+1/N}}\, \mathbf y$ with $\mathbf y \sim \mathcal{CN} (\mathbf{0}, \boldsymbol{\Sigma})$, as previously.

Hence, one can determine the false-alarm relationship:
\begin{align*}
PFA_{AMF}{}_{\hat{\boldsymbol{\Sigma}},\hat{\boldsymbol{\mu}}} &= \mathbb{P} \left(\Lambda_{AMF}^{(N)} {}_{\hat{\boldsymbol{\Sigma}},\hat{\boldsymbol{\mu}}}>\lambda | H_0\right) \\ &=  \mathbb{P} \left({\cfrac{(N+1)}{(N-1)}}  \Lambda_{AMF}^{(N-1)} {}_{\hat{\boldsymbol{\Sigma}}}>\lambda | H_0\right)  \\ &=\mathbb{P} (\Lambda_{AMF}^{(N-1)} {}_{\hat{\boldsymbol{\Sigma}}}>\lambda' | H_0)
\end{align*}
where $\lambda' = {\frac{(N-1)}{(N+1)}} \lambda$, which leads to the conclusion.
\end{IEEEproof}

\subsection{Kelly Detector}
The Kelly detector for both unknown mean vector and covariance matrix has now to be derived since it is not the previous Kelly in which an estimate of the mean is plugged. Following the same lines as in \cite{kelly1986adaptive}, we now assume that both the mean vector and the covariance matrix are unknown. The likelihood functions under $H_0$ and $H_1$ are given in \eqref{LF-Kelly}. Under $H_0$ and $H_1$, the maxima are achieved at
\begin{equation*}
\max_{\boldsymbol{\Sigma},\boldsymbol{\mu}} f_i = \left(\cfrac{1}{(\pi e)^m|\mathbf T_i|}\right)^{N+1}, \text{ for } i =0,1,
\end{equation*}
where
\begin{multline*}
(N+1)\mathbf{T}_0 = (\mathbf{x} - {\boldsymbol{\mu}_0})(\mathbf{x} - {\boldsymbol{\mu}_0})^H + \sum_{i=1}^N   (\mathbf{x}_i - {\boldsymbol{\mu}_0})(\mathbf{x}_i - {\boldsymbol{\mu}_0})^H,\\
(N+1)\mathbf{T}_1 =(\mathbf{x} - \alpha \mathbf{p}-{\boldsymbol{\mu}_1}) (\mathbf{x} - \alpha \mathbf{p}-{\boldsymbol{\mu}_1})^H+ \sum_{i=1}^N   (\mathbf{x}_i - {\boldsymbol{\mu}_1})(\mathbf{x}_i - {\boldsymbol{\mu}_1})^H,
\end{multline*}
and
\begin{align}\label{mu0}
{\boldsymbol{\mu}_0} & = \cfrac{1}{N+1}\left( \mathbf x + \sum_{i=1}^N\mathbf x_i\right),\\
{\boldsymbol{\mu}_1} & = \cfrac{1}{N+1}\left( \mathbf x - \alpha \mathbf p + \sum_{i=1}^N\mathbf x_i\right).\label{mu1}
\end{align}
And neglecting the exponent $N+1$, one obtains the following LR:
\begin{equation*}
L(\alpha) = \cfrac{|\mathbf T_0|}{|\mathbf T_1|}  \underset {H_0} {\overset {H_1} {\gtrless}} \eta
\end{equation*}
Then, as this LR still depends on the unknown amplitude $\alpha$ of the signal, thus, it has to be maximized w.r.t $\alpha$, which is equivalent to minimize $\mathbf T_1$ w.r.t $\alpha$. A way to do this is to introduce the following sample covariance matrix:
\begin{equation}\label{S0}
\mathbf S_0 = \sum_{i=1}^N (\mathbf x_i - \boldsymbol{\mu}_0)(\mathbf x_i - \boldsymbol{\mu}_0)^H.
\end{equation}
Then, $(N+1)|\mathbf T_0|$ can be written as
\begin{equation*}
(N+1)|\mathbf T_0| = |\mathbf S_0| \left(1+(\mathbf{x} - {\boldsymbol{\mu}_0})^H\,\mathbf S_0^{-1}\,(\mathbf{x} - {\boldsymbol{\mu}_0})\right).
\end{equation*}
In the same way, and after some manipulations, $(N+1)|\mathbf T_1|$ becomes
\begin{align*}
(N+1)|\mathbf T_1| = & |\mathbf S_0| \left(\sum_{i=1}^N (\mathbf{x}_i - {\boldsymbol{\mu}_1})^H\,\mathbf S_0^{-1}\,(\mathbf{x}_i - {\boldsymbol{\mu}_1}) \right. \\ & + \left.  \vphantom{\sum_{i=1}^N}(\mathbf{x} - \alpha \mathbf p - {\boldsymbol{\mu}_1})^H\mathbf S_0^{-1}(\mathbf{x} - \alpha \mathbf p -{\boldsymbol{\mu}_1})\right)\\ & = |\mathbf S_0| (A+B).
\end{align*}
Now, let us rewrite the two terms $A$ and $B$ to separate the terms involving $\alpha$. By recalling that ${\boldsymbol{\mu}_1}={\boldsymbol{\mu}_0} - \cfrac{1}{N+1}\,\alpha\,\mathbf p$, one obtains:
\begin{align*}
A = & 1+ \cfrac{N |\alpha|^2}{(N+1)^2} \mathbf p^H\mathbf S_0^{-1}\mathbf p \\ & + \cfrac{2}{N+1}\,\Re\left\{\bar{\alpha} \mathbf p^H \mathbf S_0^{-1} \sum_{i=1}^N (\mathbf{x}_i - {\boldsymbol{\mu}_0})\right\},\\ B = & (\mathbf{x} - {\boldsymbol{\mu}_0})^H\,\mathbf S_0^{-1}\,(\mathbf{x} - {\boldsymbol{\mu}_0}) + \cfrac{N^2 |\alpha|^2}{(N+1)^2} \mathbf p^H\mathbf S_0^{-1}\mathbf p \\ & - \cfrac{2N}{N+1}\,\Re\left\{\bar{\alpha} \mathbf p^H \mathbf S_0^{-1} (\mathbf{x} - {\boldsymbol{\mu}_0})\right\}.
\end{align*}
Notice that $\mathbf{x} - {\boldsymbol{\mu}_0} = - \sum_{i=1}^N (\mathbf{x}_i - {\boldsymbol{\mu}_0})$, then  rearranging the expression of $(N+1)|\mathbf T_1|$ leads to
\begin{align*}
\cfrac{(N+1)|\mathbf T_1|}{ |\mathbf S_0|} = & \cfrac{(N+1)|\mathbf T_0|}{ |\mathbf S_0|} + \cfrac{N |\alpha|^2}{(N+1)} \mathbf p^H\mathbf S_0^{-1}\mathbf p \\ &- 2\,\Re\left\{\bar{\alpha} \mathbf p^H \mathbf S_0^{-1} (\mathbf{x} - {\boldsymbol{\mu}_0})\right\}.
\end{align*}
Now, the term depending on $\alpha$ can be rewritten as follows
\begin{align*}
\cfrac{N}{(N+1)} \mathbf p^H\mathbf S_0^{-1}\mathbf p  \left| \alpha - \cfrac{N+1}{N} \,\cfrac{\mathbf p^H\mathbf S_0^{-1} (\mathbf{x} - {\boldsymbol{\mu}_0})}{\mathbf p^H\mathbf S_0^{-1}\mathbf p}\right|^2 \\ - \cfrac{N+1}{N}\, \cfrac{\left|\mathbf p^H \mathbf S_0^{-1} (\mathbf{x} - {\boldsymbol{\mu}_0)}\right|^2}{\mathbf p^H\mathbf S_0^{-1}\mathbf p}.
\end{align*}
Minimizing $|\mathbf T_1|$ w.r.t $\alpha$ is equivalent to cancel the square term in the previous equation. Thus, the GLRT can now be written according to the following definition.
\begin{Def}[The generalized Kelly detector]
Under Gaussian assumptions, the extension of the Kelly's test when both the mean vector and the covariance matrix of the background are unknown takes the following form:
\begin{align}\label{Kelly-gene}
\Lambda = \cfrac{\beta(N)\,\left|\mathbf p^H \mathbf S_0^{-1} (\mathbf{x} - {\boldsymbol{\mu}_0})\right|^2}{(\mathbf p^H\mathbf S_0^{-1}\mathbf p)\left(1+(\mathbf{x} - {\boldsymbol{\mu}_0})^H\,\mathbf S_0^{-1}\,(\mathbf{x} - {\boldsymbol{\mu}_0})\right)} \underset {H_0} {\overset {H_1} {\gtrless}} \lambda,
\end{align}
where $\beta(N) = \cfrac{N+1}{N}$, $\lambda = \cfrac{\eta-1}{\eta}$ and 
\begin{itemize}
\item $\displaystyle \mathbf S_0 = \sum_{i=1}^N (\mathbf x_i - \boldsymbol{\mu}_0)(\mathbf x_i - \boldsymbol{\mu}_0)^H,$ \item $\displaystyle {\boldsymbol{\mu}_0} = \cfrac{1}{N+1}\left( \mathbf x + \sum_{i=1}^N\mathbf x_i\right).$
\end{itemize}
\end{Def}
Let us now comment on this new detector. One can notice that both the covariance matrix $\mathbf S_0$ as well as the mean ${\boldsymbol{\mu}_0}$ estimates depend on the data $\mathbf x$ under test, which is not the case in other classical detectors where the unknown parameters are estimated from signal-free secondary data. Consequently, $\mathbf S_0$ and $\mathbf{x} - {\boldsymbol{\mu}_0}$ are not independent. Moreover, the covariance matrix estimate $\mathbf S_0$ is not Wishart-distributed due to the non-standard mean estimate ${\boldsymbol{\mu}_0}$. Thus, the derivation of this ratio distribution is very difficult. \\ 

As for previous detector, it would be intuitive to think that the proposed test behaves as the classical Kelly's test but for $N-1$ degrees of freedom. To prove that let us first rewrite \eqref{Kelly-gene} as follows:
\begin{align*}
\Lambda = \cfrac{\left|\mathbf p^H \mathbf S_0^{-1} \mathbf{y}\right|^2}{(\mathbf p^H\mathbf S_0^{-1}\mathbf p)\left(1+\cfrac{N}{N+1}\,\mathbf{y}^H\,\mathbf S_0^{-1}\,\mathbf{y}\right)} \underset {H_0} {\overset {H_1} {\gtrless}} \lambda
\end{align*}
where we use:
\begin{itemize}
\item $(\mathbf{x} - {\boldsymbol{\mu}_0}) = \cfrac{N}{N+1}\,(\mathbf{x} - \hat{\boldsymbol{\mu}}_{SMV})$, \item $\hat{\boldsymbol{\mu}}_{SMV} = 1/N \sum_{i=1}^N \mathbf x_i$, \item $\mathbf{y}=\sqrt {\displaystyle\frac{N}{N+1}}\,(\mathbf{x} - \hat{\boldsymbol{\mu}}_{SMV})\sim \mathcal{CN} (\mathbf{0},\boldsymbol{\Sigma})$.
\end{itemize}

Now, let us set $\mathbf S_0^{(i)} = \sum_{i=1}^N (\mathbf x_i - \boldsymbol{\mu}_0^{(i)})(\mathbf x_i - \boldsymbol{\mu}_0^{(i)})^H$ where $\boldsymbol{\mu}_0^{(i)}=1/N(\sum_{j\neq i}^N \mathbf x_j+ \mathbf x)$. Then, the test becomes
\begin{align*}
\cfrac{\cfrac{N+1}{N}\,\left|\mathbf p^H (\mathbf S_0^{(i)})^{-1} (\mathbf{x} - \hat{\boldsymbol{\mu}}_{SMV})\right|^2}{(\mathbf p^H(\mathbf S_0^{(i)})^{-1}\mathbf p)\left(1+(\mathbf{x} - \hat{\boldsymbol{\mu}}_{SMV})^H\,(\mathbf S_0^{(i)})^{-1}\,(\mathbf{x} - \hat{\boldsymbol{\mu}}_{SMV})\right)}.
\end{align*}
One can notice that each $\mathbf x_i$ (including $\mathbf x$) plays the same role, thus the distribution of this test is the same for every permutation of the $(N+1)$-sample $(\mathbf x, \mathbf x_1, \hdots, \mathbf x_N)$. However, the dependency between the covariance matrix estimate and the data under test $\mathbf x$ still remains.\\

To fill this gap, another way of taking advantage of the Kelly's detector when the mean vector is unknown can be to use the classical scheme recalled in \eqref{LR-Kelly} and to plug the classical estimator of the mean, based only on the secondary data, i.e. $\hat{\boldsymbol{\mu}}_{SMV} = 1/N \sum_{i=1}^N \mathbf x_i$. This leads to the the plug-in Kelly's detector:
\begin{align*}
&\Lambda_{Kelly}^{(N)} {}_{\hat{\boldsymbol{\Sigma}}, \hat{\boldsymbol{\mu}}} = \\&\cfrac {|\mathbf{p}^H \, \hat{\boldsymbol{\Sigma}}^{-1}_{SCM} \, (\mathbf{x} - {\hat{\boldsymbol{\mu}}}_{SMV})|^2}{\left(\mathbf{p}^H \, \hat{\boldsymbol{\Sigma}}^{-1}_{SCM} \mathbf{p}\right) \, \left(  N +  (\mathbf{x} - {\hat{\boldsymbol{\mu}}}_{SMV})^H \, \hat{\boldsymbol{\Sigma}}^{-1}_{SCM} \, (\mathbf{x} - {\hat{\boldsymbol{\mu}}}_{SMV})\right)} \underset {H_0} {\overset {H_1} {\gtrless}} \lambda\,.
\end{align*}

In this case, the distribution can be derived. This is the purpose of the following proposition.
\begin{Prop}\label{Prop-Pfa_Kelly_mean}
The theoretical relationship between the $PFA$ and the threshold is given by
 \begin{align}\label{Pfa_Kelly_mean}
&PFA_{Kelly}{}_{\hat{\boldsymbol{\Sigma}}, \hat{\boldsymbol{\mu}}} =  \cfrac{\Gamma(N)}{\Gamma(N-m+1)\,\Gamma(m-1)} \, \nonumber \\ & \times \int_0^1 \left[1 + \frac{\lambda}{1-\lambda} \left(1-\cfrac{u}{N+1} \right)\right]^{m-N}  u^{N-m}  (1-u)^{m-2} \, du 
\end{align} 

\end{Prop}

\begin{IEEEproof}
The detection test rewritten with $\hat{\boldsymbol{\Sigma}}^{-1}_{SCM} = N \, \hat{\mathbf{W}^{-1}_{N-1}} $ becomes: 
\begin{equation*}
 \Lambda^{(N)}_{Kelly \, \hat{\boldsymbol{\Sigma}}, \hat{\boldsymbol{\mu}}} = \cfrac{N^2 \,  \left|\mathbf{p}^H\hat{\mathbf{W}}_{N-1}^{-1}(\mathbf{x} - \hat{{\boldsymbol{\mu}}})\right|^2}{N  \, \left(\mathbf{p}^H\,\hat{\mathbf{W}_{N-1}^{-1}}\,\mathbf{p}\right)\, \left(N +  N \,\mathbf{y}^H\,\hat{\mathbf{W}_{N-1}^{-1}}\,\left(\mathbf{x} - \hat{{\boldsymbol{\mu}}}\right) \right)} 
\end{equation*}
and replacing $(\mathbf{x} - {\hat{\boldsymbol{\mu}}})$ by $\sqrt {\displaystyle\frac{N+1}{N}}\,\mathbf{y}$, one obtains:
\begin{align*}
 &\Lambda^{(N)}_{Kelly \, \hat{\boldsymbol{\Sigma}}, \hat{\boldsymbol{\mu}}} = \\ =& \cfrac{\cfrac{N+1}{N}\,N^2 \,\left|\mathbf{p}^H\,\hat{\mathbf{W}}_{N-1}^{-1}\,\mathbf{y}\right|^2}{N \, \left(\mathbf{p}^H\,\hat{\mathbf{W}_{N-1}^{-1}}\,\mathbf{p}\right)\left(N + \cfrac{N+1}{N} \,N \,\mathbf{y}^H\,\hat{\mathbf{W}_{N-1}^{-1}}\,\mathbf{y}\right)} \\
= & \cfrac{\left|\mathbf{p}^H\,\hat{\mathbf{W}}_{N-1}^{-1}\,\mathbf{y}\right|^2}{ \left(\mathbf{p}^H\,\hat{\mathbf{W}_{N-1}^{-1}}\,\mathbf{p}\right)\left(\cfrac{N}{N+1} + \mathbf{y}^H\,\hat{\mathbf{W}_{N-1}^{-1}}\,\mathbf{y}\right)}
\label{kelly_wishart}
\end{align*}
with $\mathbf{y}\sim \mathcal{CN} (\mathbf{0},\boldsymbol{\Sigma})$. \\
The classical Kelly detector obtained when the mean vector is known is recalled here, built with $N-1$ zero-mean Gaussian data, and written with $\hat{\mathbf{W}}_{N-1}$:
 \begin{equation}
\Lambda^{(N-1)}_{Kelly \, \hat{\boldsymbol{\Sigma}}} =  \cfrac{\left|\mathbf{p}^H\,\hat{\mathbf{W}}_{N-1}^{-1}\,\mathbf{y}\right|^2}{ \left(\mathbf{p}^H\,\hat{\mathbf{W}_{N-1}^{-1}}\,\mathbf{p}\right)\left(1 + \mathbf{y}^H\,\hat{\mathbf{W}_{N-1}^{-1}}\,\mathbf{y}\right)}
\label{Kellynmoinsun}
\end{equation}
It is worth pointing out that the term ${N}/{(N+1)}$ resulting from the mean estimation in $\Lambda^{(N)}_{Kelly \, \hat{\boldsymbol{\Sigma}}, \hat{\boldsymbol{\mu}}}$ does not appear in the classical Kelly detector \eqref{Kellynmoinsun}. This fact prevents from relating the two expressions. Thus, a proof similar to the Proposition \ref{Prop-Pfa_AMF_mean} is not feasible.\\
According to \cite{Kraut01,richmond2000performance}, an equivalent LR can be expressed as:
\begin{equation*}
\widehat{\kappa}^2 = \displaystyle \frac{\Lambda_{Kelly \, \hat{\boldsymbol{\Sigma}}, \hat{\boldsymbol{\mu}}}^{(N)}}{1- \Lambda_{Kelly \, \hat{\boldsymbol{\Sigma}}, \hat{\boldsymbol{\mu}}}^{(N)}}
\underset {H_0} {\overset {H_1} {\gtrless}} \displaystyle \frac{\lambda}{1-\lambda}
\end{equation*}
Following the same development proposed in \cite{Kraut01}, the statistic $\widehat{\kappa}^2$ can be identified as the ratio $\theta / \beta$ between two independent scalar random variables  $\theta$ and $\beta$. For this particular development of Kelly distribution with non-centered data, the scalar random variable $\beta$ is found to have the same distribution as the function $1-u/(N+1)$ where $u$ is a random variable following a complex central beta distribution with $N-m+1,m-1$ degrees of freedom:
\begin{equation*}
u \sim f_u(u) =  \cfrac{\Gamma(N)}{\Gamma(N-m+1)\,\Gamma(m-1)} \,u^{N-m} \, (1-u)^{m-2}
\end{equation*}
whereas the p.d.f. of the variable $\theta$ is distributed according to the complex $F$-distribution with $1, N-m$ degrees of freedom scaled by $1/(N-m)$:
 \begin{equation*}
\theta \sim f_{\theta}(\theta) = (N-m) \, (1+\theta)^{m-N-1}
\end{equation*} 
One can now derive the cumulative density function of the Kelly test as:
\begin{align*}
&\mathbb{P}\left( \Lambda_{Kelly \, \hat{\boldsymbol{\Sigma}}, \hat{\boldsymbol{\mu}}}^{(N)} \leq \lambda\right)  = \mathbb{P}\left( \widehat{\kappa}^2 \leq \, \displaystyle \frac{\lambda}{1-\lambda} \right) = \mathbb{P}\left( \theta \leq \beta \, \displaystyle \frac{\lambda}{1-\lambda} \right) \\&= \int_0^1 \left[\int_0^{\frac{\lambda}{1-\lambda}\, \left(1-u/(N+1)\right)} f_{\theta}(v)\, dv \right] \, f_u(u)  \, du
\end{align*}
Solving the integral one obtains the "PFA-threshold" relationship:
 \begin{align*}
&PFA_{Kelly}{}_{\hat{\boldsymbol{\Sigma}}, \hat{\boldsymbol{\mu}}} =  \cfrac{\Gamma(N)}{\Gamma(N-m+1)\,\Gamma(m-1)} \, \\ & \times \int_0^1 \left[1 + \frac{\lambda}{1-\lambda} \left(1-\cfrac{u}{N+1} \right)\right]^{m-N}  u^{N-m}  (1-u)^{m-2} \, du 
\label{Kellyrel}
\end{align*} 
However, the final expression can not be further simplified and a closed-form expression as those obtained for the other detectors can not be determined.
\end{IEEEproof}
\subsection{Adaptive Normalized Matched Filter}
Similarly, the ANMF for both mean vector and covariance matrix estimation becomes:
\begin{equation*}
\Lambda_{ANMF} {}_{\hat{\boldsymbol{\Sigma}}, \hat{\boldsymbol{\mu}}} = \cfrac {|\mathbf{p}^H \, {\hat{\boldsymbol{\Sigma}}}^{-1} \, (\mathbf{x} - \hat{{\boldsymbol{\mu}}})|^2}{(\mathbf{p}^H \, {\hat{\boldsymbol{\Sigma}}}^{-1}\mathbf{p}) \, \left( (\mathbf{x} - {\hat{\boldsymbol{\mu}}})^H \, {\hat{\boldsymbol{\Sigma}}}^{-1} \, (\mathbf{x} - {\hat{\boldsymbol{\mu}}})\right)} \underset {H_0} {\overset {H_1} {\gtrless}} \lambda\,.
\end{equation*}

\begin{Prop}\label{Prop-Pfa_ANMF_mean}
The theoretical relationship between the $PFA$ and the threshold is given by
\begin{equation}\label{Pfa_ANMF_mean}
PFA_{ANMF}{}_{\hat{\boldsymbol{\Sigma}},\hat{\boldsymbol{\mu}}} = (1-\lambda)^{a-1} {}_2F_1\left(a,a-1;b-1;\lambda \right)\, ,
\end{equation}
where $a = (N-1)-m+2$ and $b = (N-1)+2$. 
\end{Prop}

\begin{IEEEproof}
The proof is similar to the proof of Proposition \ref{Prop-Pfa_AMF_mean}. The main difference is due to the normalization term $(\mathbf{x} - {\hat{\boldsymbol{\mu}}})^H \, {\hat{\boldsymbol{\Sigma}}}^{-1} \, (\mathbf{x} - {\hat{\boldsymbol{\mu}}})$. Indeed, the correction factor $N/(N-1)$ appears both at the numerator and at the denominator, and consequently, it disappears. The same argument is also true for the factor $N$ that arises from the covariance matrix estimates, i.e. since the detector is homogeneous in terms of covariance matrix estimates, this scalar also disappears. Thus, the distribution of the ANMF with an estimate of the mean is exactly the same as in eq. \eqref{pfa_ANMF} where  $N$ is replaced by $N-1$.
\end{IEEEproof}
\section{Simulations}
\label{sec4}
\input{fig-AMF-1.tex}

In this section, we validate the theoretical analysis on simulated data.
The experiments were conducted on $m=5$ dimensional Gaussian vectors, for different values of $N$, the number of secondary data and the computations have been made through $10^6$ Monte-Carlo trials. The true covariance is chosen as a Toeplitz matrix whose entries are $\Sigma_{i,j} = \rho^{|i-j|}$ and where $\rho = 0.4$. The mean vector is arbitrary set to have all entries equal to $(3+4j)$.

\subsection{False Alarm Regulation}
The FA regulation is presented for previous detection schemes having a closed-form expression, i.e. for all except the generalized Kelly detector. Fig. \ref{AMF} shows the false-alarm regulation for the MF, the AMF when only covariance matrix is unknown and the AMF for both covariance matrix and mean vector unknown. The perfect agreement of the green and yellow curves illustrates the results of Proposition \ref{Prop-Pfa_AMF_mean}. Moreover, remark that when $N$ increases both AMF get closer to each other, and they approach the known parameters case MF.

Fig. \ref{Kelly-Gauss} and Fig. \ref{ANMF-Gauss} present the FA regulation for the Kelly detector and the ANMF respectively, under Gaussian assumption. 
For clarity purposes, the results are displayed in terms of the threshold $\eta$ from \eqref{LR-Kelly}, $\eta = (1-\lambda)^{-(N+1)}$,
 and \eqref{LR-ANMF},  $\eta = (1-\lambda)^{-m}$, respectively and a logarithmic scale is used. This validates results of Proposition \ref{Prop-Pfa_Kelly_mean} and \ref{Prop-Pfa_ANMF_mean} for the SCM-SMV. \\

\input{fig-Kelly-1.tex}
\input{fig-ANMF-Gauss2-1.tex}

Remark that the derived relationships given by eqs. \eqref{Pfa_AMF_mean} and \eqref{Pfa_ANMF_mean} are quite similar to those for which the mean is known. However, as illustrated in Fig. \ref{AMF} and Fig.\ref{ANMF-Gauss},  there is an important difference for small values of $N$.
It is worth pointing out that the theoretical "PFA-threshold" relationships presented above depend only on the size of the vectors $m$ and the number of secondary data used to estimate the parameters $N$. Thus, the detector outcome will not depend on the true value of the covariance matrix or the mean vector. These three detectors hold the CFAR property with respect to the background parameters. However, their distribution strongly relies on the underlying distribution of the background, ie. if Gaussian assumption is not fulfilled the  "PFA-threshold" relationship will divert from the theoretical results derived in this paper.

\subsection{Performance Evaluation}
\begin{figure}[!htp]
\centering
\begin{tikzpicture}[font=\footnotesize,scale=1]
\renewcommand{\axisdefaulttryminticks}{8}
\pgfplotsset{every major grid/.append style={dashed}}
\pgfplotsset{every axis legend/.append style={fill=white,cells={anchor=west},at={(0.02,0.98)},anchor=north west}}
\tikzstyle{dashed dotted}=[dash pattern=on 1pt off 5pt on 6pt off 2pt]

\begin{axis}[xlabel=SNR dBs, ylabel=$P_d$
,grid=major,
xmin=-0,xmax=30,ymin=0,ymax=1,
grid=major,
ylabel style={yshift=-5pt},
]
\addplot[green, dashed, smooth,line width=1.5pt] plot coordinates {
(-10,0.0008)(-8.97727,0.001)(-7.95455,0.001)(-6.93182,0.001)(-5.90909,0.0013)(-4.88636,0.0009)(-3.86364,0.0013)(-2.84091,0.0023)(-1.81818,0.0017)(-0.795455,0.0019)(0.227273,0.0019)(1.25,0.0036)(2.27273,0.0035)(3.29545,0.0054)(4.31818,0.0061)(5.34091,0.0099)(6.36364,0.0133)(7.38636,0.0199)(8.40909,0.0326)(9.43182,0.05)(10.4545,0.0778)(11.4773,0.1203)(12.5,0.1916)(13.5227,0.2874)(14.5455,0.4245)(15.5682,0.5865)(16.5909,0.7431)(17.6136,0.8798)(18.6364,0.9543)(19.6591,0.9899)(20.6818,0.9981)(21.7045,0.9998)(22.7273,1)(23.75,1)(24.7727,1)(25.7955,1)(26.8182,1)(27.8409,1)(28.8636,1)(29.8864,1)(30.9091,1)(31.9318,1)(32.9545,1)(33.9773,1)(35,1)
};
\addplot[red, dashed,smooth,line width=1pt] plot coordinates {
(-10,0.001)(-8.97727,0.0012)(-7.95455,0.0014)(-6.93182,0.0009)(-5.90909,0.0015)(-4.88636,0.0018)(-3.86364,0.0023)(-2.84091,0.0034)(-1.81818,0.0026)(-0.795455,0.0037)(0.227273,0.004)(1.25,0.0064)(2.27273,0.0073)(3.29545,0.0107)(4.31818,0.0102)(5.34091,0.016)(6.36364,0.0233)(7.38636,0.0328)(8.40909,0.0469)(9.43182,0.0712)(10.4545,0.1011)(11.4773,0.1467)(12.5,0.2038)(13.5227,0.2784)(14.5455,0.3591)(15.5682,0.4709)(16.5909,0.5742)(17.6136,0.6785)(18.6364,0.7712)(19.6591,0.8553)(20.6818,0.9084)(21.7045,0.9469)(22.7273,0.9747)(23.75,0.9881)(24.7727,0.9954)(25.7955,0.9983)(26.8182,0.9997)(27.8409,0.9995)(28.8636,0.9998)(29.8864,1)(30.9091,1)(31.9318,1)(32.9545,1)(33.9773,1)(35,1)
};
\addplot[blue, dashed, smooth,line width=1.5pt] plot coordinates {
(-10,0.001)(-8.97727,0.001)(-7.95455,0.0012)(-6.93182,0.0012)(-5.90909,0.0014)(-4.88636,0.0017)(-3.86364,0.0016)(-2.84091,0.0029)(-1.81818,0.0029)(-0.795455,0.0026)(0.227273,0.0033)(1.25,0.0053)(2.27273,0.0063)(3.29545,0.0093)(4.31818,0.0101)(5.34091,0.0169)(6.36364,0.0241)(7.38636,0.0382)(8.40909,0.0574)(9.43182,0.089)(10.4545,0.1364)(11.4773,0.2036)(12.5,0.3029)(13.5227,0.4172)(14.5455,0.5505)(15.5682,0.704)(16.5909,0.8176)(17.6136,0.9095)(18.6364,0.9574)(19.6591,0.9843)(20.6818,0.9944)(21.7045,0.9972)(22.7273,0.9994)(23.75,0.9998)(24.7727,1)(25.7955,1)(26.8182,1)(27.8409,1)(28.8636,1)(29.8864,1)(30.9091,1)(31.9318,1)(32.9545,1)(33.9773,1)(35,1)
};

\addplot[black, dotted, smooth,line width=1.5pt] plot coordinates {
(-10,0.0009)(-8.97727,0.0012)(-7.95455,0.0012)(-6.93182,0.0015)(-5.90909,0.0016)(-4.88636,0.0017)(-3.86364,0.0018)(-2.84091,0.0029)(-1.81818,0.0031)(-0.795455,0.0031)(0.227273,0.0039)(1.25,0.0058)(2.27273,0.0075)(3.29545,0.0113)(4.31818,0.0112)(5.34091,0.0196)(6.36364,0.028)(7.38636,0.0424)(8.40909,0.0626)(9.43182,0.097)(10.4545,0.1489)(11.4773,0.2184)(12.5,0.3145)(13.5227,0.4232)(14.5455,0.5471)(15.5682,0.6844)(16.5909,0.785)(17.6136,0.868)(18.6364,0.9173)(19.6591,0.951)(20.6818,0.9695)(21.7045,0.9796)(22.7273,0.9865)(23.75,0.9914)(24.7727,0.9936)(25.7955,0.9947)(26.8182,0.9956)(27.8409,0.9956)(28.8636,0.9975)(29.8864,0.997)(30.9091,0.9976)(31.9318,0.9973)(32.9545,0.9974)(33.9773,0.9975)(35,0.9983)
};

\legend{\scriptsize{AMF }, \scriptsize{ANMF},  \scriptsize{"Plug-in" Kelly}, \scriptsize{Generalized Kelly}};
\end{axis}
\end{tikzpicture}
\caption{Probability of detection for different SNR values and $PFA=10^{-3}$ in Gaussian case.}
\label{SNR_simu}
\end{figure}
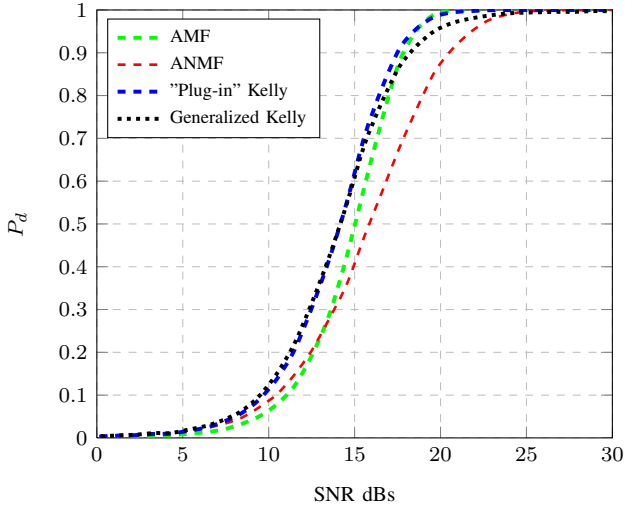
The four detection schemes are compared in terms of probability of detection. Firstly, one sets the probability of false alarm to an specific value. Here we set $PFA = 10^{-3}$ with $m = 5$ and $N=10$. Then, the threshold is adjusted to reach the desired PFA, according to the false alarm regulation curves described above. For the generalized Kelly detector, the threshold is empirically computed to ensure the same $PFA = 10^{-3}$.
Fig. \ref{SNR_simu} presents the detection probability versus the SNR. When data follow a multivariate normal distribution, the detectors delivering the best performance results are the Kelly detectors ("Plug-in" and generalized). Actually, these detectors lead to very similar performance with a small improvement of the generalized (resp. "plug in") one at low (resp. high) SNR. As expected, the AMF and the ANMF require a higher SNR to achieve same performance. \\

\subsection{Hyperspectral Real Data}

\begin{figure}[htb]
\centering
\includegraphics[width=0.3\textwidth]{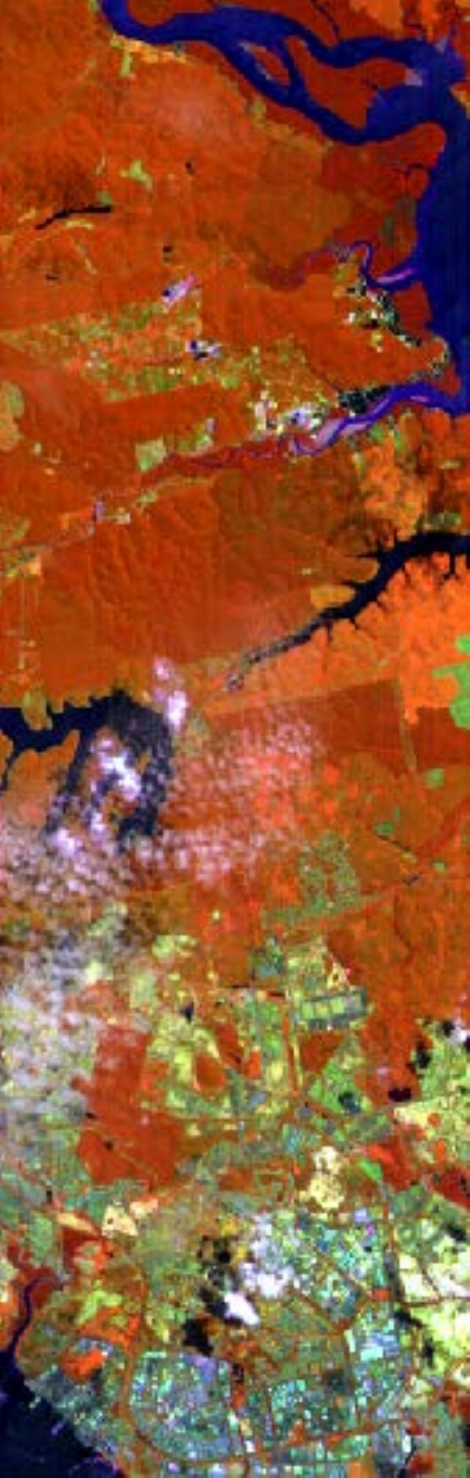}
\caption{True color composition of the Hyperion scene.}
 \label{fig:image}
\end{figure}
\begin{figure}
\centering
\begin{tikzpicture}[font=\footnotesize,scale=1]
\renewcommand{\axisdefaulttryminticks}{8}
\pgfplotsset{every major grid/.append style={dashed}}
\pgfplotsset{every axis legend/.append style={fill=white,cells={anchor=west},at={(0.02,0.98)},anchor=north west}}
\tikzstyle{dashed dotted}=[dash pattern=on 1pt off 5pt on 6pt off 2pt]

\begin{axis}[xlabel=Normal Quantiles, ylabel= Quantiles of Input Samples
,grid=major,
xmin=-3.5,xmax=3,ymin=-0.01,ymax=0.025,
grid=major,
ylabel style={yshift=-5pt},
]
\addplot[only marks, blue,mark=x] plot coordinates {
(-3.341479e+00,-8.389421e-03)(-2.376031e+00,-4.697469e-03)(-2.118099e+00,-3.354206e-03)(-1.952884e+00,-2.298647e-03)(-1.828330e+00,-1.387186e-03)(-1.727006e+00,-9.401895e-04)(-1.640827e+00,-4.794431e-04)(-1.565354e+00,-3.558501e-17)(-1.497871e+00,4.216525e-04)(-1.436594e+00,5.292863e-04)(-1.380281e+00,1.008132e-03)(-1.328035e+00,1.515919e-03)(-1.279181e+00,1.899475e-03)(-1.233203e+00,2.269745e-03)(-1.189694e+00,2.442849e-03)(-1.148328e+00,2.753017e-03)(-1.108838e+00,2.958797e-03)(-1.071006e+00,3.163282e-03)(-1.034648e+00,3.314330e-03)(-9.996085e-01,3.462119e-03)(-9.657552e-01,3.716129e-03)(-9.329741e-01,3.817893e-03)(-9.011661e-01,4.071734e-03)(-8.702447e-01,4.241180e-03)(-8.401339e-01,4.505144e-03)(-8.107661e-01,4.617245e-03)(-7.820815e-01,4.738248e-03)(-7.540265e-01,4.995897e-03)(-7.265527e-01,5.145191e-03)(-6.996168e-01,5.313386e-03)(-6.731791e-01,5.574349e-03)(-6.472039e-01,5.716013e-03)(-6.216582e-01,5.943112e-03)(-5.965119e-01,6.049359e-03)(-5.717372e-01,6.178777e-03)(-5.473087e-01,6.399322e-03)(-5.232025e-01,6.515665e-03)(-4.993966e-01,6.649660e-03)(-4.758705e-01,6.773544e-03)(-4.526048e-01,6.871277e-03)(-4.295816e-01,6.997733e-03)(-4.067840e-01,7.192362e-03)(-3.841958e-01,7.293564e-03)(-3.618020e-01,7.410310e-03)(-3.395882e-01,7.559978e-03)(-3.175407e-01,7.696509e-03)(-2.956466e-01,7.855909e-03)(-2.738932e-01,8.005491e-03)(-2.522688e-01,8.126947e-03)(-2.307616e-01,8.238049e-03)(-2.093607e-01,8.388553e-03)(-1.880553e-01,8.524736e-03)(-1.668348e-01,8.701168e-03)(-1.456893e-01,8.947012e-03)(-1.246087e-01,9.142675e-03)(-1.035834e-01,9.227399e-03)(-8.260369e-02,9.334880e-03)(-6.166033e-02,9.443526e-03)(-4.074398e-02,9.535586e-03)(-1.984544e-02,9.706972e-03)(1.044429e-03,9.915060e-03)(2.193476e-02,1.002793e-02)(4.283466e-02,1.012098e-02)(6.375330e-02,1.023485e-02)(8.469987e-02,1.038968e-02)(1.056837e-01,1.053377e-02)(1.267141e-01,1.060740e-02)(1.478008e-01,1.074422e-02)(1.689534e-01,1.084079e-02)(1.901818e-01,1.093886e-02)(2.114963e-01,1.103389e-02)(2.329074e-01,1.112687e-02)(2.544257e-01,1.123700e-02)(2.760626e-01,1.135954e-02)(2.978295e-01,1.150905e-02)(3.197384e-01,1.166794e-02)(3.418019e-01,1.177870e-02)(3.640330e-01,1.189261e-02)(3.864456e-01,1.212911e-02)(4.090540e-01,1.227912e-02)(4.318735e-01,1.243632e-02)(4.549202e-01,1.255170e-02)(4.782110e-01,1.290505e-02)(5.017643e-01,1.303422e-02)(5.255993e-01,1.312720e-02)(5.497367e-01,1.329729e-02)(5.741987e-01,1.346998e-02)(5.990093e-01,1.359793e-02)(6.241943e-01,1.376846e-02)(6.497816e-01,1.392988e-02)(6.758015e-01,1.406578e-02)(7.022873e-01,1.423295e-02)(7.292752e-01,1.444564e-02)(7.568051e-01,1.463932e-02)(7.849208e-01,1.483174e-02)(8.136712e-01,1.510439e-02)(8.431105e-01,1.530686e-02)(8.732992e-01,1.550880e-02)(9.043057e-01,1.576710e-02)(9.362069e-01,1.590389e-02)(9.690906e-01,1.617003e-02)(1.003057e+00,1.637158e-02)(1.038222e+00,1.658219e-02)(1.074720e+00,1.683503e-02)(1.112709e+00,1.708960e-02)(1.152376e+00,1.729239e-02)(1.193944e+00,1.748494e-02)(1.237684e+00,1.770605e-02)(1.283929e+00,1.793256e-02)(1.333097e+00,1.832926e-02)(1.385717e+00,1.845752e-02)(1.442481e+00,1.878082e-02)(1.504316e+00,1.919176e-02)(1.572506e+00,1.942380e-02)(1.648907e+00,1.977046e-02)(1.736361e+00,1.999105e-02)(1.839557e+00,2.057410e-02)(1.967144e+00,2.109051e-02)(2.138206e+00,2.264068e-02)(2.412735e+00,2.412828e-02)
};
\addplot[red, dashed,smooth,line width=1pt] plot coordinates {
(-3.341479e+00,-1.123574e-02)(3.341479e+00,3.086108e-02)
};

%
\end{axis}
\end{tikzpicture}
\caption{Q-Q Plot of the data sample versus the Normal theoretical distribution.}
\label{fig:qqplot}
\end{figure}
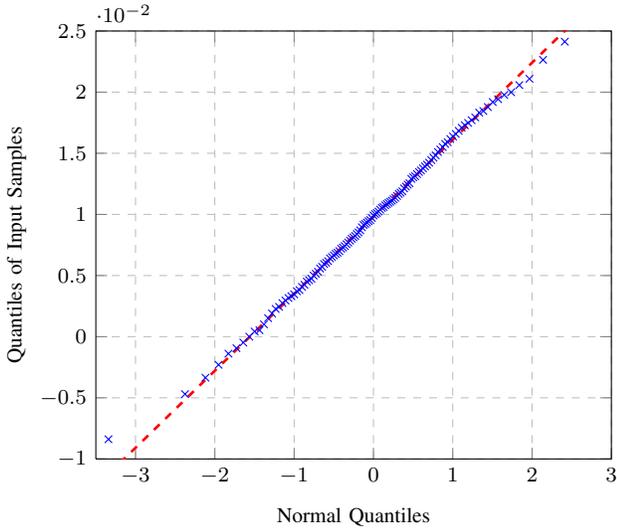

The same experiments have been conducted on a real hyperspectral image. The scene analyzed is the NASA Hyperion sensor dataset displayed in Fig. \ref{fig:image}. The image is constituted of $798 \times 253$ pixels and 116 spectral bands after water absorption bands have been removed. The analysis has been done on a homogenous part of the image corresponding to the water region on the top left of the image. The part extracted consists on $ 60 \times 20$ pixels. In order to ensure the validity of the proposed methods, we show in Fig. \ref{fig:qqplot} the outcome of a classical Gaussianity test "Q-Q plot" for the selected region over the band 42. However, these techniques allow to "validate" the Gaussianity of each band but cannot ensure the Gaussianity of the corresponding vector.\\

\input{fig-AMF-HSI2.tex}

Since hyperspectral data are real and positive, we propose to use a Hilbert filter in order to render them complex. A downsampling taking one over two consecutive bands is required to avoid redundant information that can reduce the covariance matrix rank. However, it is important to note that the real component after Hilbert transform is still the original signal. 
To avoid the well-known problem due to high dimensionality, we have chosen sequentially six bands in the complex representation. In this approach, both covariance matrix and mean vector are estimated using a sliding window of size $5\times 5$, having $ N = 24$ secondary data. \\

The outcome of the detectors for this image are shown on the Fig. \ref{AMF-HSI}, Fig.  \ref{Kelly-HSI} and Fig. \ref{ANMF-HSI} respectively. The results obtained on real HSI data on a  Gaussian distributed region agree with the theoretical relationships presented above. Remark that the false-alarm rate that can be achieved depends on the number of points on which the detector is calculated (in a similar manner to the Monte-Carlo trials). As the homogenous area is bounded and the data set is small, the distribution of the detectors may divert for small values of the PFA directly related to the size of the region.\\
\input{fig-Kelly-HSI2.tex}
\input{fig-ANMF-HSI2.tex}

Depending on the underlying material, the distribution of the detector might divert from the expected behavior when Gaussian distribution is assumed. This is the case on these real data since the extracted area is not perfectly Gaussian. This suggests the use of non-Gaussian distributions to model the background for hyperspectral imaging. 


\section{Conclusion}
\label{sec5}

Four adaptive detection schemes, the AMF,  Kelly detectors with a "plug-in" and a generalized versions as well as the ANMF, have been analyzed in the case where both the covariance matrix and the mean vector are unknown and need to be estimated. In this context, theoretical closed-form expressions for false-alarm regulation have been derived under Gaussian assumptions for the SCM-SMV estimates for three detection schemes. The resulting "PFA-threshold" expressions highlight the CFARness of these detectors since they only depend on the size and the number of data, but not on the unknown parameters. The theoretical analysis has been validated through Monte Carlo simulations and the performances of the detectors have been compared in terms of probability of detection. Finally, the analysis on experimental hyperspectral data validates the theoretical contribution through real application, in which a homogeneous subset of data has been extracted.
But more generally, this work finds its purpose in signal processing methods for which both mean vector and covariance matrix are unknown.

\appendices
\section{Complex Normal distributions}
\label{app1}
A $m$-dimensional vector $\mathbf{x} = \mathbf{u} + j \mathbf{v}$ has a complex normal distribution with mean $\boldsymbol{\mu}$ and covariance matrix $\boldsymbol{\Sigma} = E[(\mathbf{x} - \boldsymbol{\mu})(\mathbf{x} - \boldsymbol{\mu})^H]$, denoted $\mathcal{CN}(\boldsymbol{\mu},\boldsymbol{\Sigma})$, if $\mathbf{z} = (\mathbf{u}^T, \mathbf{v}^T)^T \in \mathbb{R}^{2m}$ has a normal distribution \cite{van1995multivariate}. If $\text{rank}(\boldsymbol{\Sigma}) = m$, the probability density function exists and is of the form
\[
f_{\mathbf{x}} (\mathbf{x}) = \pi^{-m} |\boldsymbol{\Sigma}|^{-1} \exp \{-(\mathbf{x}- \boldsymbol{\mu})^H \boldsymbol{\Sigma}^{-1}(\mathbf{x}- \boldsymbol{\mu})\}.
\]
The resulting Maximum Likelihood Estimates (MLE) are the well-known SCM and SMV defined as:
\[
\hat{\boldsymbol{\mu}}_{SMV} =  \cfrac{1}{N}\sum_{i=1}^N \mathbf{x}_i \quad \hat{\boldsymbol{\Sigma}}_{SCM} =  \cfrac{1}{N}\sum_{i=1}^N (\mathbf{x}_i-\hat{\boldsymbol{\mu}}) (\mathbf{x}_i-\hat{\boldsymbol{\mu}})^H
\]
where the $\mathbf{x}_i$ are independent and identically distributed (IID) $\mathcal{CN}(\boldsymbol{\mu},\boldsymbol{\Sigma})$.

\section{Wishart distribution}
\label{app2}
Let $\mathbf{x}_1,...,\mathbf{x}_N$ be an IID $N$-sample, where $\mathbf{x}_i \sim \mathcal{CN} (\boldsymbol{\mu},\boldsymbol{\Sigma})$. Let us define $\hat{\boldsymbol{\mu}} = \hat{\boldsymbol{\mu}}_{SMV}$ and $\hat{\mathbf{W}} = N \, \hat{\boldsymbol{\Sigma}}_{SCM}$ referred to as a Wishart matrix. Thus one has (see \cite{Gupta00} for the real case):
\begin{itemize}
\item $\hat{\boldsymbol{\mu}}$ and $\hat{\mathbf{W}}$ are independently distributed;
\item $\hat{\boldsymbol{\mu}} \sim \mathcal{CN} (\boldsymbol{\mu}, \frac{1}{N}\boldsymbol{\Sigma})$; 
\item $\hat{\mathbf{W}} \sim \mathcal{CW}(N-1, \boldsymbol{\Sigma})$ is Whishart distributed with $N-1$ degrees of freedom
\end{itemize}





\bibliographystyle{IEEEtran}

\bibliography{refs}

\begin{thebibliography}{10}
\providecommand{\url}[1]{#1}
\csname url@samestyle\endcsname
\providecommand{\newblock}{\relax}
\providecommand{\bibinfo}[2]{#2}
\providecommand{\BIBentrySTDinterwordspacing}{\spaceskip=0pt\relax}
\providecommand{\BIBentryALTinterwordstretchfactor}{4}
\providecommand{\BIBentryALTinterwordspacing}{\spaceskip=\fontdimen2\font plus
\BIBentryALTinterwordstretchfactor\fontdimen3\font minus
  \fontdimen4\font\relax}
\providecommand{\BIBforeignlanguage}[2]{{%
\expandafter\ifx\csname l@#1\endcsname\relax
\typeout{** WARNING: IEEEtran.bst: No hyphenation pattern has been}%
\typeout{** loaded for the language `#1'. Using the pattern for}%
\typeout{** the default language instead.}%
\else
\language=\csname l@#1\endcsname
\fi
#2}}
\providecommand{\BIBdecl}{\relax}
\BIBdecl

\bibitem{chang2003hyperspectral}
C.-I. Chang, \emph{Hyperspectral imaging: techniques for spectral detection and
  classification}.\hskip 1em plus 0.5em minus 0.4em\relax Springer, 2003,
  vol.~1.

\bibitem{manolakis2002detection}
D.~Manolakis and G.~Shaw, ``Detection algorithms for hyperspectral imaging
  applications,'' \emph{Signal Processing Magazine, IEEE}, vol.~19, no.~1, pp.
  29--43, 2002.

\bibitem{stein2002anomaly}
D.~W. Stein, S.~G. Beaven, L.~E. Hoff, E.~M. Winter, A.~P. Schaum, and A.~D.
  Stocker, ``Anomaly detection from hyperspectral imagery,'' \emph{Signal
  Processing Magazine, IEEE}, vol.~19, no.~1, pp. 58--69, 2002.

\bibitem{chang2002anomaly}
C.-I. Chang and S.-S. Chiang, ``Anomaly detection and classification for
  hyperspectral imagery,'' \emph{Geoscience and Remote Sensing, IEEE
  Transactions on}, vol.~40, no.~6, pp. 1314--1325, 2002.

\bibitem{kwon2006kernel}
H.~Kwon and N.~M. Nasrabadi, ``Kernel matched subspace detectors for
  hyperspectral target detection,'' \emph{Pattern Analysis and Machine
  Intelligence, IEEE Transactions on}, vol.~28, no.~2, pp. 178--194, 2006.

\bibitem{matteoli2010tutorial}
S.~Matteoli, M.~Diani, and G.~Corsini, ``A tutorial overview of anomaly
  detection in hyperspectral images,'' \emph{Aerospace and Electronic Systems
  Magazine, IEEE}, vol.~25, no.~7, pp. 5--28, 2010.

\bibitem{kay1998fundamentals}
S.~M. Kay, \emph{Fundamentals of Statistical signal processing, Volume 2:
  Detection theory}.\hskip 1em plus 0.5em minus 0.4em\relax Prentice Hall PTR,
  1998.

\bibitem{gini2001selected}
F.~Gini, A.~Farina, and M.~Greco, ``Selected list of references on radar signal
  processing,'' \emph{Aerospace and Electronic Systems, IEEE Transactions on},
  vol.~37, no.~1, pp. 329--359, 2001.

\bibitem{manolakis2003hyperspectral}
D.~Manolakis, D.~Marden, and G.~Shaw, ``Hyperspectral image processing for
  automatic target detection applications,'' \emph{Lincoln Laboratory Journal},
  vol.~14, no.~1, pp. 79--116, 2003.

\bibitem{reed1990adaptive}
I.~Reed and X.~Yu, ``Adaptive multiple-band cfar detection of an optical
  pattern with unknown spectral distribution,'' \emph{Acoustics, Speech and
  Signal Processing, IEEE Transactions on}, vol.~38, no.~10, pp. 1760--1770,
  1990.

\bibitem{mahalanobis1936generalized}
P.~C. Mahalanobis, ``On the generalized distance in statistics,''
  \emph{Proceedings of the National Institute of Sciences (Calcutta)}, vol.~2,
  pp. 49--55, 1936.

\bibitem{kelly1986adaptive}
E.~J. Kelly, ``An adaptive detection algorithm,'' \emph{Aerospace and
  Electronic Systems, IEEE Transactions on}, no.~2, pp. 115--127, 1986.

\bibitem{Kraut01}
S.~Kraut, L.~L. Scharf, and L.~T. Mc~Whorter, ``{Adaptive Subspace
  Detectors},'' \emph{Signal Processing, IEEE Transactions on}, vol.~49, no.~1,
  pp. 1--16, January 2001.

\bibitem{robey1992cfar}
F.~C. Robey, D.~R. Fuhrmann, E.~J. Kelly, and R.~Nitzberg, ``A cfar adaptive
  matched filter detector,'' \emph{Aerospace and Electronic Systems, IEEE
  Transactions on}, vol.~28, no.~1, pp. 208--216, 1992.

\bibitem{kraut1999cfar}
S.~Kraut and L.~L. Scharf, ``{The CFAR adaptive subspace detector is a
  scale-invariant GLRT},'' \emph{Signal Processing, IEEE Transactions on},
  vol.~47, no.~9, pp. 2538--2541, 1999.

\bibitem{frontera2013false}
J.~Frontera-Pons, F.~Pascal, and J.~Ovarlez, ``False-alarm regulation for
  target detection in hyperspectral imaging,'' in \emph{Computational Advances
  in Multi-Sensor Adaptive Processing (CAMSAP), 2013 IEEE 5th International
  Workshop on}.\hskip 1em plus 0.5em minus 0.4em\relax IEEE, 2013, pp.
  161--164.

\bibitem{manolakis2009there}
D.~Manolakis, R.~Lockwood, T.~Cooley, and J.~Jacobson, ``Is there a best
  hyperspectral detection algorithm?'' in \emph{SPIE Defense, Security, and
  Sensing}.\hskip 1em plus 0.5em minus 0.4em\relax International Society for
  Optics and Photonics, 2009, pp. 733\,402--733\,402.

\bibitem{manolakis2013remarkable}
D.~Manolakis, E.~Truslow, M.~Pieper, T.~Cooley, M.~Brueggeman, and S.~Lipson,
  ``The remarkable success of adaptive cosine estimator in hyperspectral target
  detection,'' in \emph{SPIE Defense, Security, and Sensing}.\hskip 1em plus
  0.5em minus 0.4em\relax International Society for Optics and Photonics, 2013,
  pp. 874\,302--874\,302.

\bibitem{Gini02}
F.~Gini and M.~V. Greco, ``Covariance matrix estimation for {CFAR} detection in
  correlated heavy tailed clutter,'' \emph{Signal Processing, special section
  on SP with Heavy Tailed Distributions}, vol.~82, no.~12, pp. 1847--1859,
  December 2002.

\bibitem{Conte02a}
E.~Conte, A.~De~Maio, and G.~Ricci, ``Recursive estimation of the covariance
  matrix of a compound-{G}aussian process and its application to adaptive
  {CFAR} detection,'' \emph{IEEE Trans.-SP}, vol.~50, no.~8, pp. 1908--1915,
  August 2002.

\bibitem{frontera2012a}
J.~Frontera-Pons, M.~Mahot, J.~Ovarlez, F.~Pascal, and J.~Chanussot, ``Robust
  detection using \emph{M}-estimators for hyperspectral imaging,'' in
  \emph{Workshop on Hyperspectral Image and Signal Processing: Evolution in
  Remote Sensing}, 2012.

\bibitem{Pascal06}
F.~Pascal, J.-P. Ovarlez, P.~Forster, and P.~Larzabal, ``On a sirv-cfar
  detector with radar experimentations in impulsive noise,'' in \emph{Proc. of
  the European Signal Processing Conf.}, Florence, September 2006.

\bibitem{abramowitz1964handbook}
M.~E. Abramowitz \emph{et~al.}, \emph{Handbook of mathematical functions: with
  formulas, graphs, and mathematical tables}.\hskip 1em plus 0.5em minus
  0.4em\relax Courier Dover Publications, 1964, vol.~55.

\bibitem{scharf1994matched}
L.~L. Scharf and B.~Friedlander, ``Matched subspace detectors,'' \emph{Signal
  Processing, IEEE Transactions on}, vol.~42, no.~8, pp. 2146--2157, 1994.

\bibitem{richmond2000performance}
C.~D. Richmond, ``Performance of a class of adaptive detection algorithms in
  nonhomogeneous environments,'' \emph{Signal Processing, IEEE Transactions
  on}, vol.~48, no.~5, pp. 1248--1262, 2000.

\bibitem{van1995multivariate}
A.~van~den Bos, ``The multivariate complex normal distribution-a
  generalization,'' \emph{Information Theory, IEEE Transactions on}, vol.~41,
  no.~2, pp. 537--539, 1995.

\bibitem{Gupta00}
A.~K. Gupta and D.~K. Nagar, \emph{Matrix Variate Distributions}.\hskip 1em
  plus 0.5em minus 0.4em\relax Chapman \& Hall/CRC, 2000.

\end{thebibliography}
%
%
%

\end{document}